\pdfoutput=1
\documentclass[journal]{IEEEtran}
\usepackage{subcaption}
\usepackage{graphicx}
\usepackage{mathtools}
\usepackage[dvipsnames]{xcolor}
\usepackage[latin1]{inputenc}
\usepackage{times,amsmath}
\usepackage{amsfonts}
\usepackage{pstool}
\usepackage{multirow}
\usepackage{multicol}
\usepackage{enumerate}
\usepackage{graphicx}
\usepackage{mathtools, cuted}
\usepackage{MnSymbol}
\usepackage{hyperref}
\usepackage{stfloats}
\usepackage{diagbox}
\usepackage{booktabs}
\usepackage{array}

\begin{document}

\title{Multi-class Network Intrusion Detection with Class Imbalance via LSTM \& SMOTE}
 
\author{
\IEEEauthorblockN{Muhammad Wasim\ Nawaz$^\dagger$, Rashid\ Munawar$^\dagger$,  Ahsan \ Mehmood$^\ast$, Muhammad\ Mahboob\ Ur\ Rahman$^{\ast}$, Qammer\ H.\ Abbasi$^\perp$  \\
$^ { \dagger}$Department of Computer Engineering, The University of Lahore, Pakistan \\
$^\ast$Electrical engineering department, Information Technology University, Lahore, Pakistan \\
$^\perp$Department of Electronics and Nano Engineering, University of Glasgow, Glasgow, G12 8QQ, UK\\
$^\dagger$Muhammad.wasim@dce.edu.pk,
$^\ast$mahboob.rahman@itu.edu.pk,
$^\perp$qammer.Abbasi@glasgow.ac.uk\\
}
}

\maketitle
\thispagestyle{plain}
\pagestyle{plain}


 
\begin{abstract}
Monitoring network traffic to maintain the quality of service (QoS) and to detect network intrusions in a timely and efficient manner is essential. As network traffic is sequential, recurrent neural networks (RNNs) such as long short-term memory (LSTM) are suitable for building network intrusion detection systems. However, in the case of a few dataset examples of the rare attack types, even these networks perform poorly. This paper proposes to use oversampling techniques along with appropriate loss functions to handle class imbalance for the detection of various types of network intrusions. Our deep learning model employs LSTM with fully connected layers to perform multi-class classification of network attacks. We enhance the representation of minority classes: i) through the application of the Synthetic Minority Over-sampling Technique (SMOTE), and ii) by employing categorical focal cross-entropy loss to apply a focal factor to down-weight examples of the majority classes and focus more on hard examples of the minority classes. Extensive experiments on KDD99 and CICIDS2017 datasets show promising results in detecting network intrusions (with many rare attack types, e.g., Probe, R2L, DDoS, PortScan, etc.).
\end{abstract}
\section{Introduction}
Network traffic monitoring is essential for effective troubleshooting in modern communication networks such as cellular networks and the Internet of Things (IoT) that generate vast amounts of traffic data. Identifying various types of attacks enables a network administrator to take timely measures to prevent the deterioration in the network performance. The diverse nature of traffic data necessitates innovative methods for monitoring and analyzing network data. 

Early detection of trends and abnormal behaviors in computer networks is vital. Intrusion detection systems (IDSs) need to swiftly identify unusual spikes, like an increase in peer-to-peer (P2P) traffic, to prevent service disruptions. Delayed detection can lead to significant financial losses \cite{lee2014network}. Studies have been conducted for various types of network traffic monitoring and analysis, which include anomaly detection \cite{alrawashdeh2016toward} and traffic classification \cite{azab2022network}.

Traditional intrusion detection systems relying on patterns and static signatures face limitations due to evolving attack types. IDSs employing data mining can efficiently analyze vast network data, making them essential for detecting intrusions and abnormalities \cite{sun2020dl}.
Furthermore, network management systems can incorporate forecasting algorithms to enhance the network's overall performance and achieve balanced resource allocation \cite{barabas2011evaluation}. 

\begin{figure}[t]
    \centering
    \includegraphics[width=.98\linewidth]{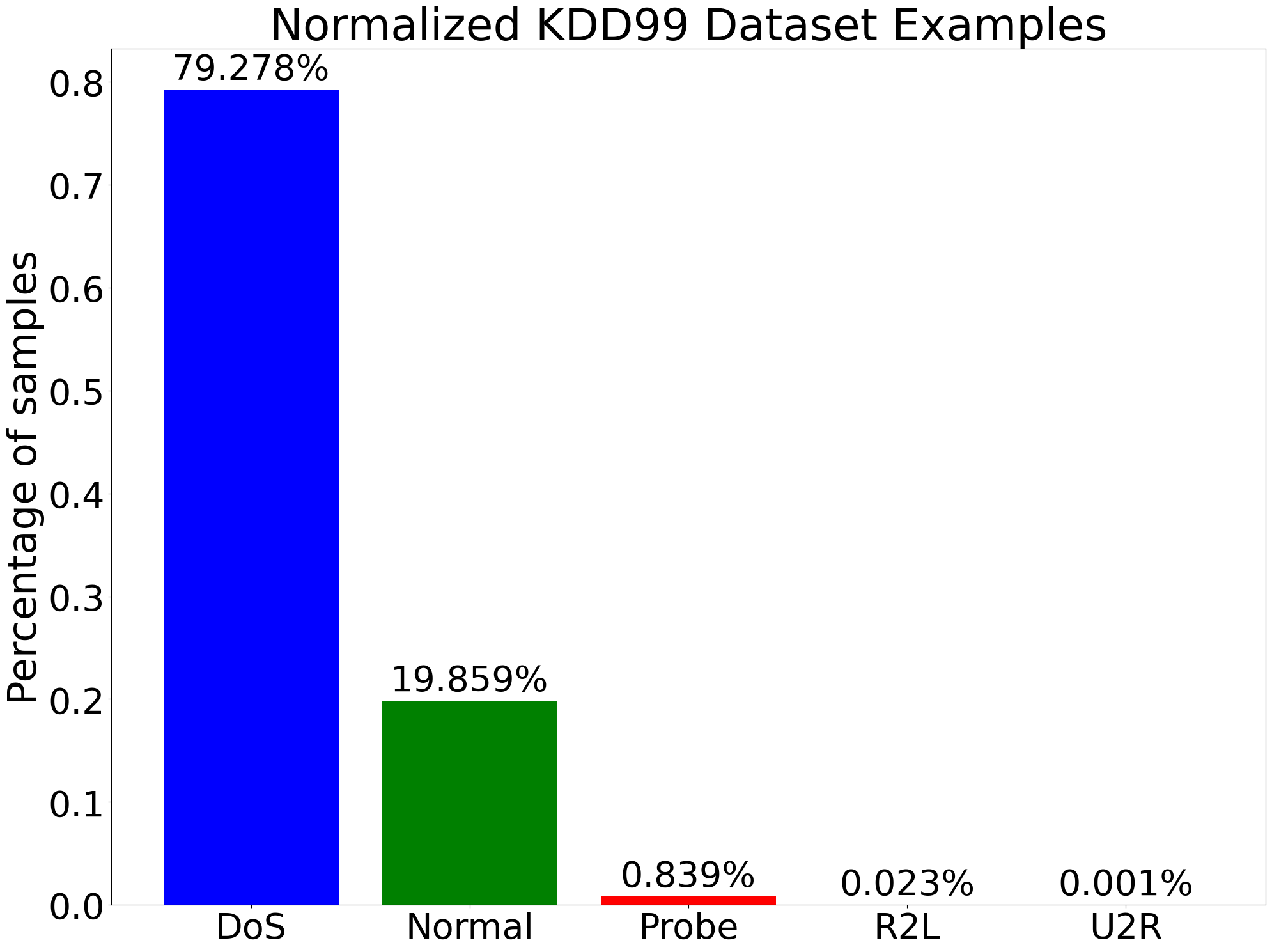}
    \caption{The number of examples of each attack type for KDD99 dataset.}
    \label{KDD99_examples}
\end{figure}

Open-source dataset KDD99 covers multiple attack scenarios such as Probe, DoS, U2R, and R2L \cite{kdd99_dataset, tavallaee2009detailed}. Another dataset, CICIDS2017, covers eight attack scenarios, including Brute Force, Heartbleed, Botnet, DoS, DDoS, and Web Attack that is helpful in the development of effective intrusion detection models \cite{sharafaldin2018toward, panigrahi2018detailed}. Both datasets are highly imbalanced as shown in figures \ref{KDD99_examples} and \ref{CICIDS2017_examples}.

{\bf Contributions:}
The main contributions of our study are given below: 
\begin{itemize}
\item Effective handling of imbalanced data in intrusion detection through the application of the Synthetic Minority Over-sampling Technique (SMOTE), enhancing the representation of minority classes.
\item Proposing a novel hybrid architecture that integrates LSTM (Long Short-Term Memory) with fully connected layers and dropout layers, demonstrating improved performance in network intrusion detection.
\item Applying Categorical Focal Loss as the loss function, providing higher emphasis on the minority classes and thus enhancing the model's ability to learn from them effectively.
\item Conducting extensive experiments on two significant datasets, CICIDS2017 and KDD99, demonstrating the model's adaptability and effectiveness across different intrusion detection contexts.
\item Outperforming existing techniques in terms of accuracy, precision, recall, and F1-score, validating the efficiency and efficacy of the proposed approach.
\item Generalization of the proposed method to effectively detect various intrusion types, including DoS, DDoS, Probe, Web Attack, Bot, PortScan, R2L, and U2R thus contributing to a comprehensive intrusion detection solution.
\end{itemize}

{\bf Outline:}
Section II provides the literature review. Section III presents the methodology. Section IV discusses the experimental setup, experiments, and the discussion of the results. Section V provides the conclusion. 

\begin{figure}[t]
    \centering
   \includegraphics[width=0.48\textwidth, height=0.3\textheight]{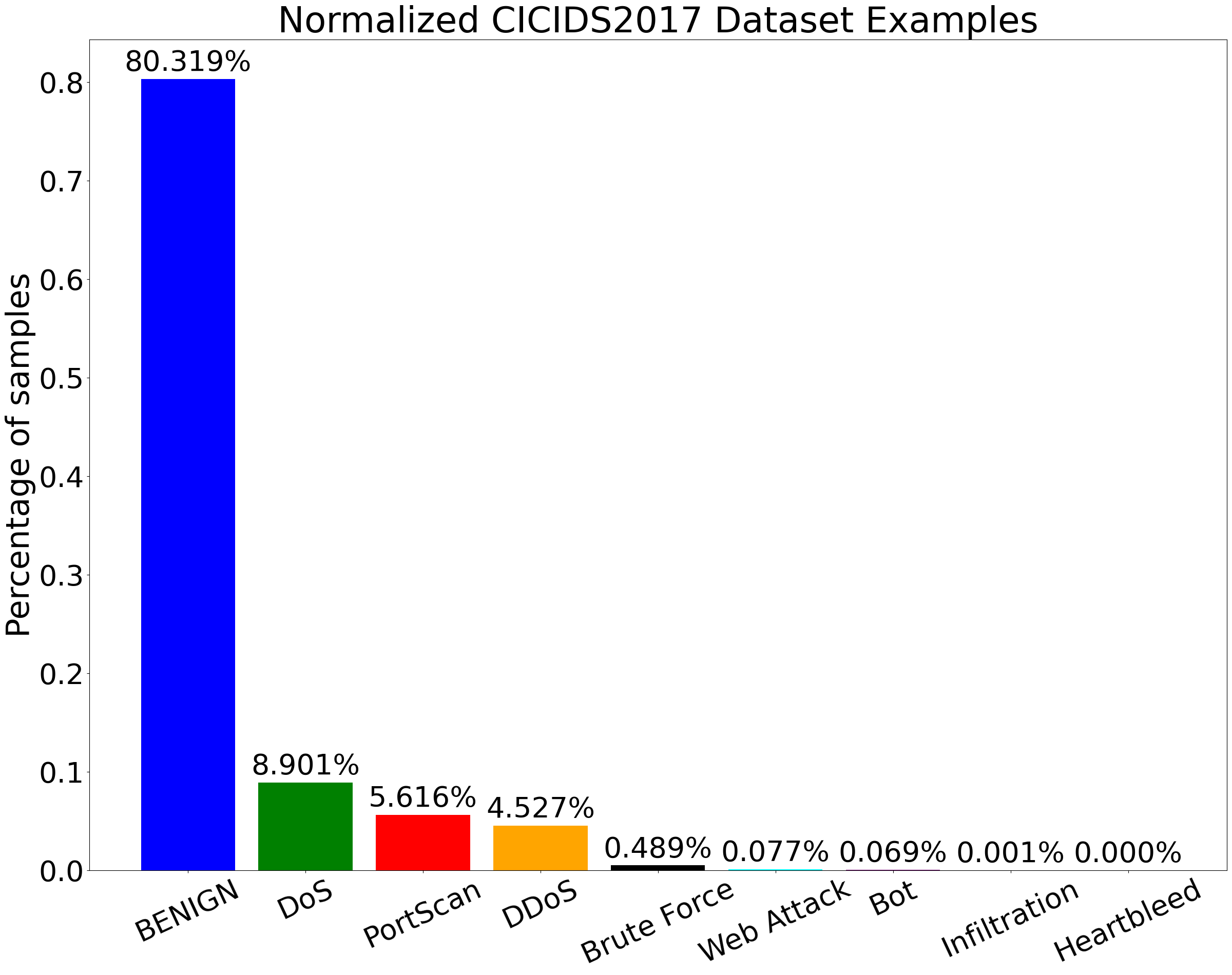}
    \caption{The number of examples of each attack type for CICIDS2017 dataset.}
    \label{CICIDS2017_examples}
\end{figure}


%
%
%
\section{Related work}
Existing techniques encompass statistical analysis, threshold-based anomaly detection, and machine learning for detecting network traffic irregularities. For intrusion detection, various approaches have been proposed. Fries et al. introduced a fuzzy-genetic method, demonstrating its applicability \cite{fries2008fuzzy}. Koc et al. utilized the Hidden Nave Bayes (HNB) model, particularly effective in identifying DoS attacks \cite{koc2012network}. In host-based intrusion detection, a two-tier architecture is commonly adopted. Zanero et al. utilized unsupervised clustering in the first tier and anomaly detection in the second tier \cite{zanero2004unsupervised}. Principal Component Analysis (PCA) combined with Support Vector Machines (SVMs) has been employed for optimal feature subset selection \cite{heba2010principal}. Additionally, studies have compared various unsupervised machine learning methods, such as K-means, KNN, and EM, for network flow classification \cite{alalousi2016preliminary}.

Furthermore, feature reduction and clustering methods have been explored. Ravale et al. proposed a hybrid strategy that combines the RBF kernel function of SVM with the K-means clustering algorithm to reduce the number of attributes for data points in the KDDCUP'99 dataset \cite{ravale2015feature}. Chen et al. introduced the fuzzy entropy weighted natural nearest neighbor (FEW-NNN) method, involving data dimension reduction and classification \cite{chen2020fuzzy}.

Deep learning-based approaches have gained traction in intrusion detection. Yu et al. proposed an IDS model based on the few-shot learning concept using CNN and DNN functions to extract key features \cite{yu2020intrusion}. Another study focused on using VAEs, FCNs, and Seq2Seq structures to detect network anomalies \cite{malaiya2018empirical}. Wang et al. proposed a model for flow prediction using deep learning and ensemble learning techniques \cite{wang2018network}. Multi-packet input units were used to compress raw traffic data, and a DL-based distributed denial of service (DDoS) detection model and defense system were proposed in an SDN environment \cite{li2018detection}. Additionally, a combined DL approach was proposed for detecting plagiarized software and malware-infected documents in IoT networks \cite{ullah2019cyber}. Various architectures, including LSTM and CNN, were utilized to enhance IDS performance \cite{jiang2020network}. Li et al. proposed a method using LSTM for HTTP Malicious Traffic Detection (HMTD) to extract temporal features and CNN to extract spatial features for HTTP malicious traffic detection on mobile networks \cite{li2019method}.

Autoencoders (AEs) have also been instrumental in improving IDS performance. Yan et al. used AE in combination with ML-based RF and SVM \cite{yan2018effective}. Seasonal deep Kalman filter network (S-DKFN) was introduced for detecting abnormal patterns using a graph representation and multi-granular seasonal information \cite{sun2022detecting}.
Moreover, Deep Belief Networks (DBNs) have shown promise in intrusion detection. DBNs, constructed by layering Restricted Boltzmann Machines (RBMs), were utilized for learning valuable features \cite{hinton2006fast}. Marir et al. proposed a distributed model based on DBN and multilayer SVM for large-scale network IDS using Apache Spark \cite{marir2018distributed}. Wei et al. suggested a DL-based model that uses DBN optimized by fusing particle swarm, fish swarm, and genetic algorithms to increase the detection accuracy of IDS \cite{wei2019optimization}.
Recurrent Neural Networks (RNNs) have gained attention for their applicability in IDS. Yu et al. developed an RNN-based IDS method for multi-class and binary classification for the NSL-KDD dataset \cite{yu2021network}. Xu et al. proposed an RNN-based IDS with GRU as the central memory, utilizing the Softmax activation and multilayer perceptron \cite{xu2018intrusion}. Additionally, LSTM has been extensively employed in extracting both spatial and temporal features from network traffic data \cite{sharafaldin2018toward}. Gwon et al. utilized LSTM with an embedding technique to propose network intrusion detection models \cite{gwon2019network}. Another study proposed an NIDS based on LSTM to recognize threats and retain long-term memory of them to prevent new attacks \cite{boukhalfa2020lstm}.

\section{The proposed network intrusion detection method}
This section outlines the proposed methodology for multiclass network intrusion detection using Long Short-Term Memory (LSTM) and Synthetic Minority Over-sampling Technique (SMOTE) \cite{chawla2002smote}.

\begin{figure}[t]
    \centering
    \includegraphics[width=.95\linewidth]{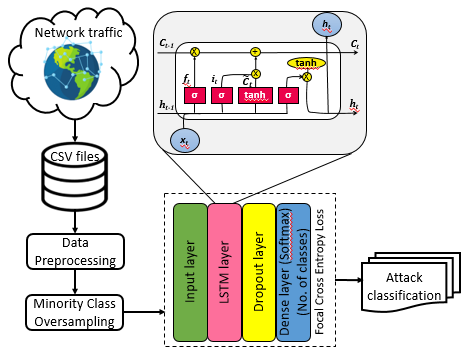}
    \caption{The proposed method.}
    \label{proposed_methodology}
\end{figure}

Figure \ref{proposed_methodology} shows the proposed method, whose main steps are summarized below:  
\begin{itemize}
    \item Preprocessing the data.
    \item Oversampling the minority classes.
    \item Constructing the LSTM-based deep learning model architecture.
    \item Choosing suitable loss function for class imbalance.
    \item Training the neural network. 
    \item Network traffic testing using the trained model.
\end{itemize}

$\textbf{Data preprocessing}$: This stage involves cleaning and preparing the data that will be used to train the neural network. This includes removing NaN (Not a Number) and Inf (Infinity) entries and standardizing the data. Furthermore, network attacks are grouped into a few main categories, simplifying the complex attack landscape and enabling targeted analysis.

\begin{figure}[t]
    \centering
    \includegraphics[width=1\linewidth]{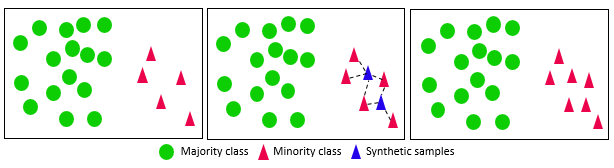}
    \caption{Samples of two classes (left), synthetic samples of the minority class are created from the existing samples (middle), by considering the neighborhood of the existing samples, and the new dataset showing the oversampled minority class (right).}
    \label{synthetic_sampling}
\end{figure}

$\textbf{SMOTE Sampling}$: To address the class imbalance, the Synthetic Minority Over-sampling Technique (SMOTE) is applied, creating synthetic instances of minority classes to get a balanced dataset. This is given in figure \ref{synthetic_sampling}, which shows samples of two classes (left), synthetic samples of the minority class are created by considering the neighborhood of the existing samples (middle). The oversampled minority class is also shown (right).

$\textbf{One Hot Label Encoder}$: The class labels are encoded into a one-hot representation.
 
$\textbf{Deep Learning Architecture}$:
The architecture of the LSTM-based neural network is designed to effectively learn and classify network traffic patterns. Our deep learning architecture leverages Long Short-Term Memory (LSTM) networks for intrusion detection. The network layers are as follows:
\begin{itemize}
\item $\textbf{Input Layer}$: Raw input data, consisting of sequences of network features, is fed into this layer. 

\item $\textbf{LSTM Layer}$: The LSTM layer processes sequences and captures temporal patterns in the network traffic data.

\item $\textbf{Dropout Layer}$: A dropout layer mitigates overfitting by randomly deactivating a portion of LSTM units during training.

\item $\textbf{Output Layer (Softmax)}$: The output layer employs the Softmax activation function, generating class probabilities for multiple attack categories. 

\item $\textbf{Categorical Focal Crossentropy Loss}$: To handle class imbalance better, the Categorical Focal Crossentropy loss (CFCL) function is utilized, giving more weight to minority classes. CFCL is an extension of the traditional cross-entropy loss function, but it assigns different weights to different classes to address the issue of class imbalance. CFCL equations are given below:
\begin{align*}
\text{CFCL}(y, \hat{y}) = -\sum_{i=1}^{N} \alpha_i (1 - \hat{y}_i)^\gamma \log(\hat{y}_i),
\end{align*}
where $N$ is the number of classes, $y$ is the true one-hot encoded class label vector, $\hat{y}$ is the predicted probability distribution over classes, $\alpha_i$ is a weighting factor for each class (typically used to address class imbalance, and $\gamma$ is the focusing parameter, which controls how much the loss focuses on hard-to-classify examples.

For each example, the loss function computes a weighted sum over all classes. The weights $\alpha_i$ are assigned to each class to account for class imbalance. Typically, classes with fewer samples receive higher weights. The term $(1 - \hat{y}_i)^\gamma$ is called the "modulating factor" that modulates the loss contribution of each class based on the predicted probability $\hat{y}_i$. If $\hat{y}_i$ is close to 1 (indicating high confidence in the prediction), this factor becomes small, reducing the loss contribution. Conversely, if $\hat{y}_i$ is close to 0 (indicating low confidence), this factor becomes large, increasing the loss contribution. The term $\log \hat{y}_i$ is the standard Cross-Entropy loss for class $i$. It measures the dissimilarity between the predicted class probabilities and the true class labels. By combining the modulating factor and the standard cross-entropy loss, the CFCL assigns higher importance to examples that are difficult to classify. This helps the model focus on improving its performance in challenging classes, effectively addressing class imbalance.

\item $\textbf{Adam Optimizer}$: The Adam optimizer is employed for efficient parameter updates during model training.
\end{itemize}
The presented methodology streamlines data preprocessing, enhances class representation, and utilizes an LSTM-based deep learning architecture tailored for network intrusion detection. This approach effectively learns complex attack patterns while mitigating class imbalance challenges.

\subsection{ Long Short Term Memory (LSTM) model}
A variety of RNN architectures are proposed for learning long-term dependencies to handle sequential data. LSTM, a variant of RNN, can effectively handle the dependency issues of long sequences due to its cell processor structure, which includes data operations for input, forget, and output gates to control the network \cite{hochreiter1997long}. 
\begin{align}
f_t &= \sigma(W_f \cdot [h_{t-1}, x_t] + b_f)  \label{lstm_eqn1} \\
i_t &= \sigma(W_i \cdot [h_{t-1}, x_t] + b_i)  \label{lstm_eqn2} \\
\tilde{C}_t &= \tanh(W_C \cdot [h_{t-1}, x_t] + b_c) \label{lstm_eqn3} \\
C_t &= f_t \odot C_{t-1} + i_t \odot \tilde{C}_t  \label{lstm_eqn4} \\
o_t &= \sigma(W_o \cdot [h_{t-1}, x_t] + b_o)  \label{lstm_eqn5} \\
h_t &= o_t \odot \tanh(C_t)  \label{lstm_eqn6} 
\end{align}
where, 

\begin{itemize}
  \item	$f_t$ is the forget gate output,
  \item	$i_t$ is the input gate output,
  \item	$\tilde{C}_t$  is the candidate cell state,
  \item   $ C_t $  is the cell state,
  \item   $ o_t$ is the output gate output,
  \item   $ h_t$ is the hidden state,
  \item   $ x_t$ is the input at time step 
  \item   $h_{t-1} $  is the hidden state from the previous time step,
  \item   $\sigma $ is the sigmoid activation function,
  \item   $\odot$ represents element-wise multiplication,
  \item  $W_f, b_f, W_i, b_i, W_C, b_C, W_o$, \text{and} $b_o$ are the weight matrices and bias terms.
\end{itemize}

The LSTM layer equations describe the forward pass of an LSTM layer over a sequence of time steps. For zero initial states:
\begin{align*}\label{eqn_lstm_states}
h_0 &= \mathbf{0} \quad \text{(initial hidden state)} \\
C_0 &= \mathbf{0} \quad \text{(initial cell state)} \\
\text{for } t &= 1, 2, \ldots, T: \\
\quad h_t, C_t &= \text{LSTM\_Cell}(h_{t-1}, C_{t-1}, x_t),
\end{align*}
where,
\begin{itemize}
  \item $h_o$ and $C_o$ are the initial hidden state and cell state,
   \item $T$ is the number of time steps,
    \item $x_t$ is the input at timestep, 
    \item $h_t$ and $C_t$ are the updated hidden state and cell state at timestep $t$,
    \item $\text{LSTM\_Cell}$ represents the LSTM cell equations \ref{lstm_eqn1} - \ref{lstm_eqn6} .
\end{itemize}

 \textbf{Forget Gate Output} ($f_t$):
It decides what to retain or forget from the previous cell state $C_{t-1}$. It's determined by a Sigmoid activation of a weighted combination of $h_{t-1}$ and $x_t$, mapping values to [0, 1]. A value of 0 means "forget completely," and 1 means "remember fully."

\textbf{Input Gate Output} ($i_t$):
It determines what new information to store in the cell state. Similar to the forget gate, it's computed using a Sigmoid activation of a weighted combination of $h_{t-1}$ and $x_t$. This activation decides which values from the candidate cell state $\tilde{C}_t$ to add to the cell state $C_t$.

\textbf{Candidate Cell State} ($\tilde{C}_t$):
It represents new information that can be added to the cell state. It's calculated by applying the hyperbolic tangent (tanh) activation to a weighted combination of $h{t-1}$ and $x_t$, squashing values to [-1, 1].

\textbf{Cell State Update} ($C_t$):
It combines the previous cell state $C_{t-1}$ and the new candidate cell state $\tilde{C}t$ based on the forget gate $f_t$ and the input gate $i_t$, using element-wise multiplication ($\odot$). If $f_t$ is close to 1, corresponding components of $C{t-1}$ are retained. If $i_t$ is close to 1, corresponding components of $\tilde{C}_t$ are added to $C_t$.

\textbf{Output Gate Output} ($o_t$):
It determines what information from the updated cell state $C_t$ should be passed to the current hidden state $h_t$. It computes a Sigmoid activation of a weighted combination of $h_{t-1}$ and $x_t$. The Sigmoid activation scales the values in $C_t$ selectively.

\textbf{Hidden State Update} ($h_t$):
It is computed as the element-wise multiplication ($\odot$) of the output gate $o_t$ and the hyperbolic tangent (tanh) of the updated cell state $C_t$. This captures relevant information passed to the next time step.

The final layer of the LSTM network is a fully connected layer with the same number of neurons as the number of attack classes. It allows us to effectively process the temporal information present in the flow data and accurately classify the different types of network attacks.

\section{ Experimental setup}
This section presents the experimental setup, evaluation metrics, and datasets for the evaluation of the proposed method. 

\subsection{Evaluation Metrics} 
To evaluate the performance of our proposed approach, we employed several standard evaluation metrics for multiclass classification: 
\begin{itemize}
    \item \textbf{Accuracy (ACC)}: The model's overall accuracy in classifying network traffic into different intrusion classes.
    $$\text{ACC} = \frac{TP + TN}{TP + TN + FP + FN}$$
    \item \textbf{Positive Predictive Value (PPV)}: The proportion of correctly classified instances for each class out of the total instances predicted as that class. The PPV is also called Precision.
    $$\text{PPV} = \frac{TP}{TP + FP}$$
    \item \textbf{True Positive Rate (TPR)}: The proportion of correctly classified instances for each class out of the total instances belonging to that class. The TPR is also called Recall.
    $$\text{TPR} = \frac{TP}{TP + FN}$$
    \item \textbf{$\bold{F_1}$-score (F1)}: The harmonic mean of precision and recall, providing a balanced measure of the model's performance for each class.
    $$\text{F1} = 2 \times \frac{PPV \times TPR}{PPV + TPR}$$
\end{itemize}
In the metrics above, TP, TN, FP, and FN are true positives, true negatives, false positives, and false negatives, respectively.
We compute these evaluation metrics on the training and test sets to assess the model's generalization ability.

\subsection{Training Setup}
During the training phase, we configured the LSTM-based deep learning model with specific parameters to optimize its performance. The following are the key aspects of our training setup:
\begin{itemize}
    \item \textbf{Batch Size}: We used a batch size of 1024, which denotes the number of samples processed in each training iteration. This value was chosen based on computational constraints and the size of our dataset.
    \item \textbf{Number of Epochs}: The model was trained for 30 epochs, indicating the number of times the entire training dataset was passed through the model during training. This value was determined through experimentation to balance training time and convergence of the model.
    \item \textbf{Learning Rate}: We employed the Adam optimizer that adaptively adjusts the learning rate during training to optimize the model's performance.
\end{itemize}
\subsection{ KDD99 Dataset}
The KDD99 dataset is a widely used benchmark dataset in the field of network intrusion detection \cite{kdd99_dataset, tavallaee2009detailed}. It was created for the purpose of evaluating intrusion detection systems' performance in detecting various types of network attacks. The dataset was generated based on network traffic data collected from a simulated environment to reflect real-world network activity. The KDD99 dataset is characterized by the following features:
\begin{itemize}
    \item $\textbf{Multiclass Classification}$: The dataset includes a diverse range of network traffic instances categorized into multiple classes based on attack types and normal network behavior.
    \item $\textbf{Attack Variety}$: The dataset covers a broad spectrum of network attacks, including Denial of Service (DoS), Probing (Probe), User-to-Root (U2R), and Remote-to-Local (R2L) attacks.
    \item $\textbf{Imbalanced Classes}$: Like many real-world scenarios, the dataset exhibits class imbalance, where certain attack types are significantly underrepresented compared to others.
    \item $\textbf{Temporal Aspect}$: The dataset captures the temporal aspects of network traffic, reflecting the sequence and timing of transmitted packets.
\end{itemize}
\begin{table}[t]
\centering
\caption{Categories of attacks of KDD99}
\begin{tabular}{ll}
\toprule
Classification  & Attack name \\
of attacks  &      \\
\midrule
Probe & Portsweep, IPsweep, Nmap, Satan \\
DoS & Neptune, Smurf, Pod, Teardrop, Land, back \\
U2R & Bufferoverflow, LoadModule, Perl, Rootkit \\
R2L & Guesspassword, Ftpwrite, Imap, Phf, Multihop, \\
 & Warezmaster, Warezclient \\
\bottomrule
\end{tabular}
\label{KDD99_attack_mapping}
\end{table}
\begin{table}[t]
\centering
\caption{Comparison of the training and test results on KDD99 dataset.}\label{KDD99_training_test}
\begin{tabular}{|c|c|c|c|c|c|}
\hline
\multirow{2}{*}{\textbf{Set}} & \multirow{2}{*}{\textbf{Class}} & \multicolumn{4}{c|}{\textbf{Metrics}} \\ 
\cline{3-6}
& & \textbf{ACC} & \textbf{PPV} & \textbf{TPR} & \textbf{F1}  \\ 
\hline
\multirow{5}{*}{\textbf{Training}} 
 & DoS & 1.00 & 1.00 & 1.00 & 1.00  \\  
 & Normal & 1.00 & 1.00 & 1.00 & 1.00  \\ 
 & Probe & 1.00 & 1.00 & 1.00 & 1.00  \\ 
 & R2L & 0.98 & 0.91 & 0.98 & 0.94 \\  
 & U2R & 0.96 & 0.88 & 0.96 & 0.92 \\ 
  \cline{2-6} 
& \textbf{Average} & \textbf{0.99} & \textbf{0.96} &\textbf{ 0.99} & \textbf{0.97} \\ 
\hline 
\multirow{5}{*}{\textbf{Test}} & DoS & 1.00 & 1.00 & 1.00 & 1.00 \\  
 & Normal & 1.00 & 0.99 & 1.00 & 0.99 \\ 
 & Probe & 0.99 & 0.98 & 0.99 & 0.98 \\ 
 & R2L & 0.99 & 0.74 & 0.98 & 0.84 \\ 
 & U2R & 0.96 & 0.52 & 0.8 & 0.63 \\ 
 \cline{2-6} 
& \textbf{Average} & \textbf{0.99} & \textbf{0.85} & \textbf{0.95} & \textbf{0.89} \\ 
\hline 
\end{tabular}
\end{table}
Since the dataset has numerous attack types that can be categorized into four main classes, we use the mapping scheme given in Table \ref{KDD99_attack_mapping}.
The training and testing results on the KDD99 dataset in Table \ref{KDD99_training_test} reveal consistent and high-performing classification across various classes using a set of metrics. During training, the model gives outstanding accuracy (ACC: 0.99) and balanced precision (PPV: 0.96) and recall (TPR: 0.99) values. This suggests the model's ability to effectively differentiate between normal and attack instances across different classes such as DoS, Probe, R2L, and U2R. In the testing phase, the model demonstrated similar effectiveness in identifying DoS and normal instances with near-perfect accuracy and precision. However, there was a slight decrease in performance, notably in distinguishing R2L and U2R attacks, reflected in lower precision and recall. Overall, the model maintains an average accuracy (ACC) of 0.99 in both training and testing, indicating its robustness in network intrusion detection.

Table \ref{KDD99_comparison_table} compares the proposed method with other methods. Support Vector Machine (SVM),  Na"ive Bayes, and J48 displayed moderate performance across all metrics. Random Forest achieved similar accuracy to SVM with slightly better precision. Conventional LSTM performed better than the aforementioned methods with an accuracy of approximately 87.3\%, highlighting the potential of recurrent neural networks. The Bidirectional LSTM \cite{imrana2018bidirectional} demonstrated notable accuracy and precision at 91.36\% and 85.81\% respectively; however, the proposed LSTM combined with SMOTE presented the highest accuracy at 98.83\%, a strong recall at 95.4\%, and the highest F1 score at 89.17\%, emphasizing its robustness in identifying intrusions. 

\begin{table}[b]
\centering
\caption{Comparison of the proposed method with other methods on KDD99 dataset.}\label{KDD99_comparison_table}
\resizebox{0.5\textwidth}{!}{%
\begin{tabular}{|m{2cm}|c|c|c|c|c|}
\hline
\textbf{Method} & \textbf{ACC} & \textbf{PPV} & \textbf{TPR} & \textbf{F1} \\
\hline
SVM & 77.12\% & 80.80\% & 77.12\% & 73.90\% \\
J48 & 75.23\% & 80.30\% & 75.23\%  & 71.3\% \\
Na\"{\i}ve Bayes & 71.48\% & 76.30\% & 71.50\%  & 71.40\% \\
Random forest & 77.07\% & 82.20\% & 77.10\% & 73.10\% \\
Conventional LSTM & 87.26\% & 90.34\% & 87.26\%  & 88.03\% \\
Bidirectional LSTM \cite{imrana2018bidirectional} & 91.36\% & \textbf{85.81\%} & 91.27\% & 88.46\% \\
LSTM+SMOTE (Proposed) & \textbf{98.83\%} & 84.61\% & \textbf{95.40\%}  & \textbf{89.17}\% \\
\hline
\end{tabular}
}
\end{table}

\begin{figure*}
    \centering
       \begin{subfigure}{0.32\linewidth} 
        \includegraphics[width=\linewidth]{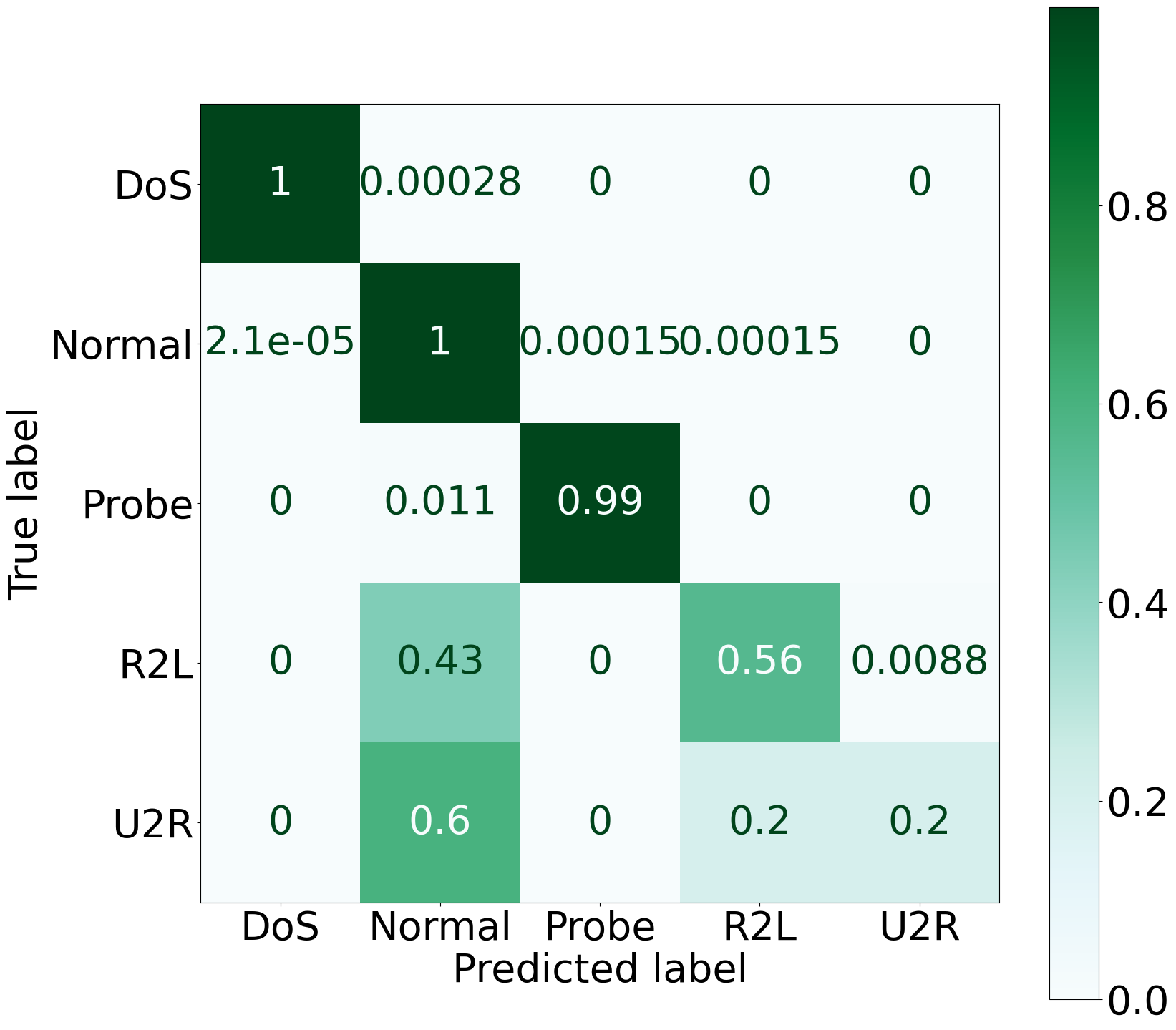}
        \caption{Using the sparse categorical cross-entropy loss, no oversampling.}
        \label{cm_kdd99_subfig1}
    \end{subfigure}
    \hfill
       \begin{subfigure}{0.32\linewidth} 
        \includegraphics[width=\linewidth]{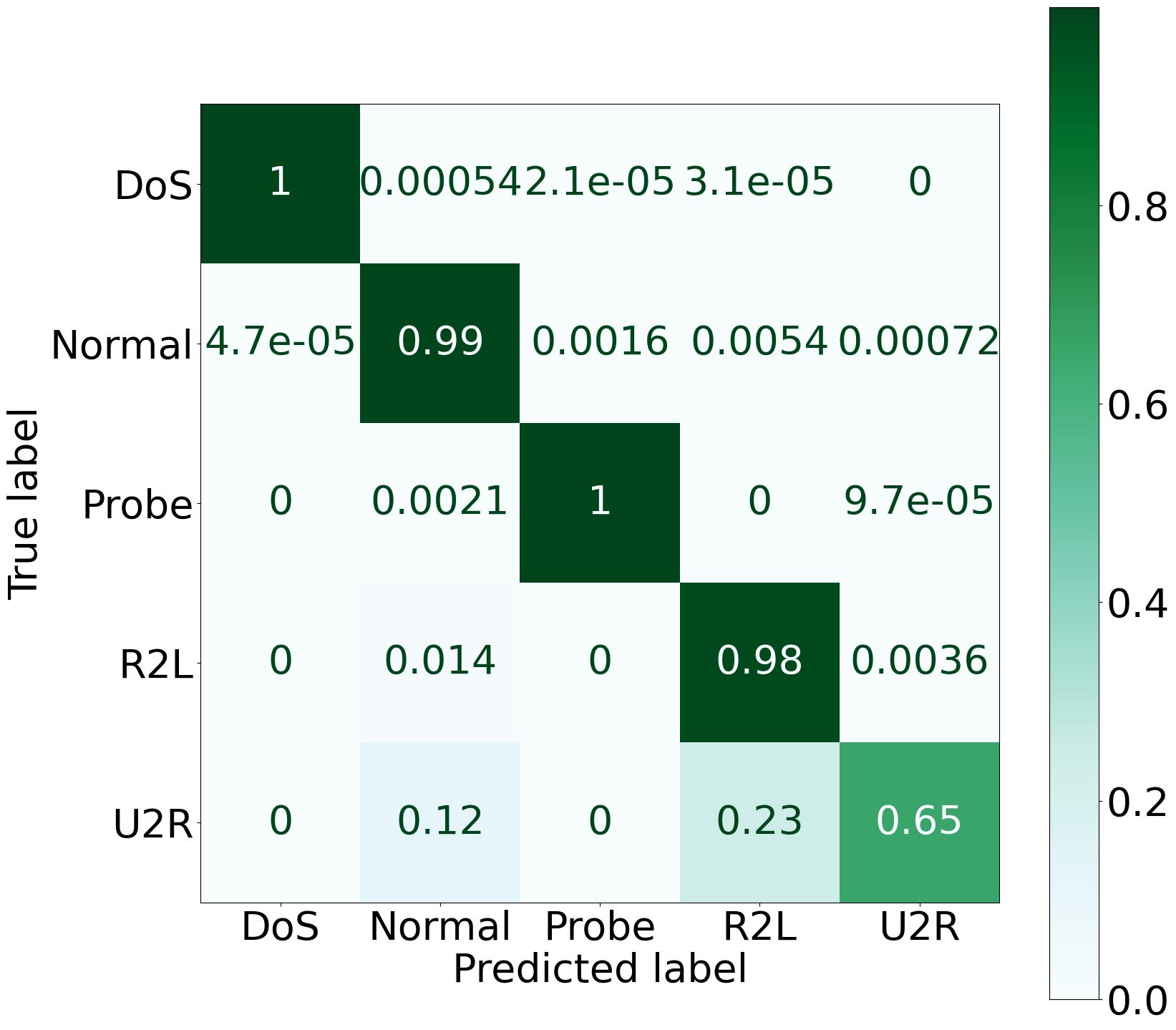}
        \caption{Using the sparse categorical cross-entropy loss with SMOTE sampling.}
        \label{cm_kdd99_subfig2}
    \end{subfigure}
    \hfill    
    \begin{subfigure}{0.32\linewidth} 
        \includegraphics[width=\linewidth]{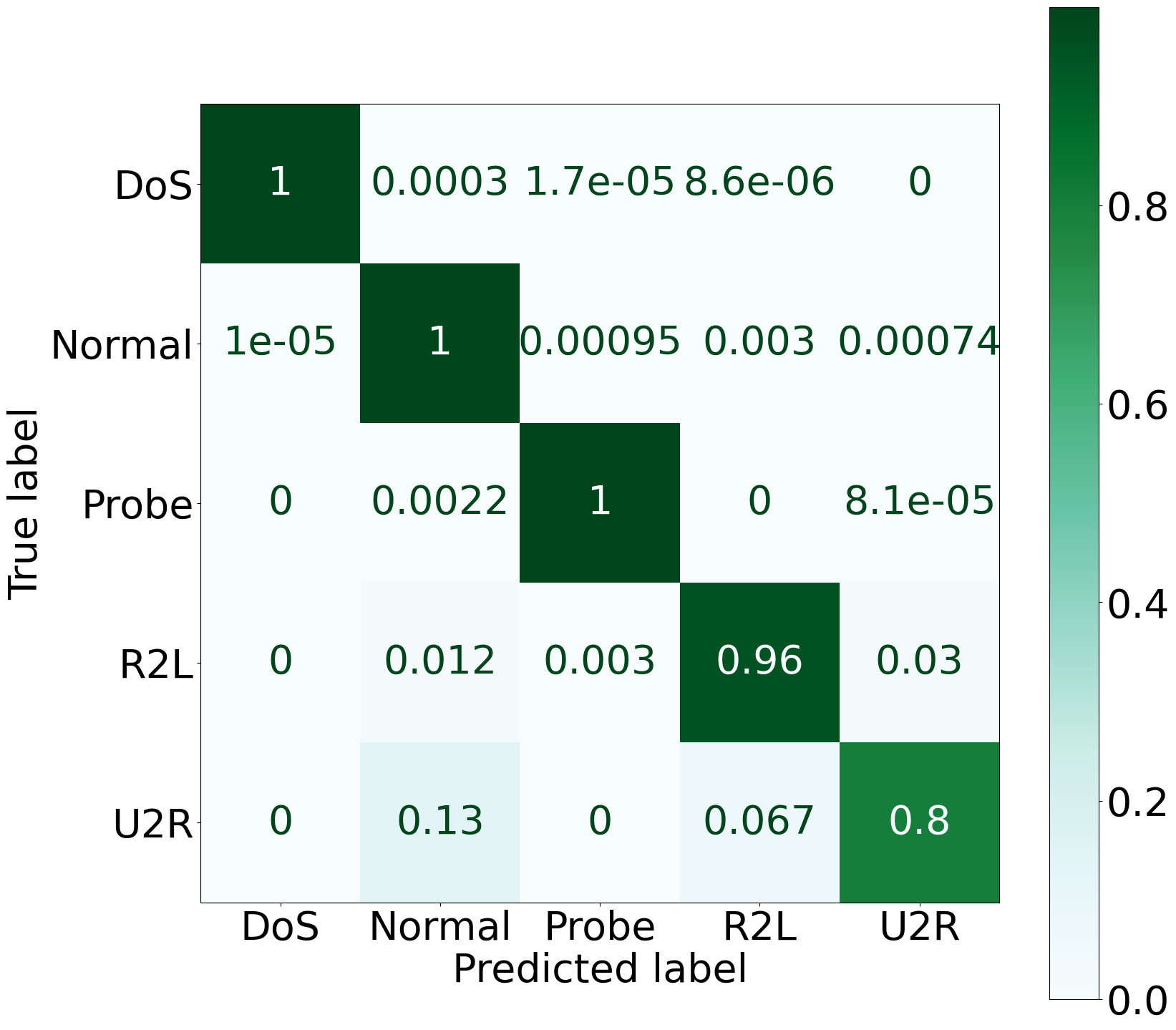}
        \caption{Using the CFCL with SMOTE sampling.}
        \label{cm_kdd99_subfig3} 
    \end{subfigure}
    \caption{Confusion matrices for KDD99 test dataset.}
    \label{cm_kdd99_fig}
\end{figure*}

%
Figure \ref{cm_kdd99_fig} shows the confusion matrices for KDD99 dataset. When the minority oversampling is not used, the results in \ref{cm_kdd99_subfig1} show the inability of the method to detect minority classes: R2L and U2R. The results in \ref{cm_kdd99_subfig2} are better; however, the presented results in \ref{cm_kdd99_subfig3} indicate the effectiveness of the proposed LSTM with SMOTE and CFCL to detect minority classes: R2L and U2R. 

\subsection{CICIDS2017 Dataset} 
We have also evaluated the performance of the proposed method on CICIDS2017 dataset \cite{sharafaldin2018toward} that provides a comprehensive representation of real-world network traffic and serves as a suitable evaluation platform for our proposed intrusion detection model. It encompasses a wide range of network traffic scenarios, including multiple types of attacks and normal traffic. Thus, the dataset allows us to assess the model's performance in detecting various network intrusions accurately. 

\begin{figure}[t]
    \centering
    \includegraphics[width=.85\linewidth, height=0.75\linewidth]{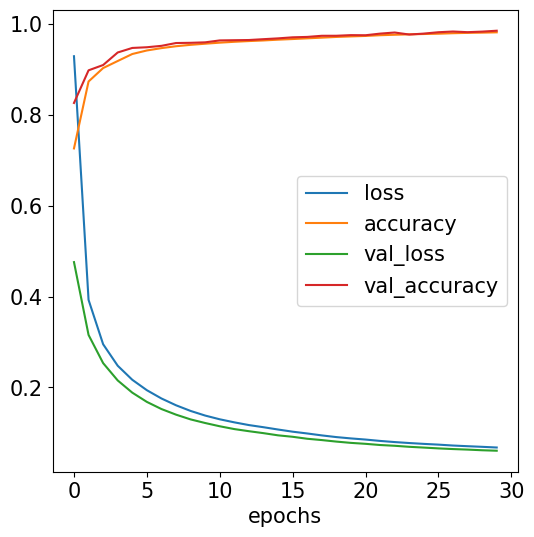}
    \caption{The loss and accuracy curves for CICIDS2017 dataset.}
    \label{CICIDS2017_loss_accuracy}
\end{figure}

The CICIDS2017 dataset consists of a total of 78 features, including traffic flow statistics, transport layer protocols, and payload-related features. It can be used for binary classification by considering all the attacks to be a single class. However, here we use it for multi-class classification. 

The CICIDS2017 dataset consists of eight files containing traffic data from five regular activity days and various cyber-attacks. We merge the traffic data from all five days into a single dataset for our experiments. To group the similar types of attacks, we use the following mapping.
\begin{itemize}
    \item DoS Hulk, DoS GoldenEye, DoS slowloris, DoS Slowhttptest: These are all variations of Denial-of-Service (DoS) attacks. We group them together under the label "\textbf{DoS}". 
    \item FTP-Patator, SSH-Patator: These are both types of Brute Force attacks. We group them under the label "\textbf{Brute Force}".
    \item Web Attack, Web Attack-XSS, Web Attack-Sql Injection: These are all types of Web Attacks. We group them together under the label "\textbf{Web Attack}".
\end{itemize}

\begin{figure}
    \centering
        \begin{subfigure}{0.84\linewidth} 
        \includegraphics[width=\linewidth]{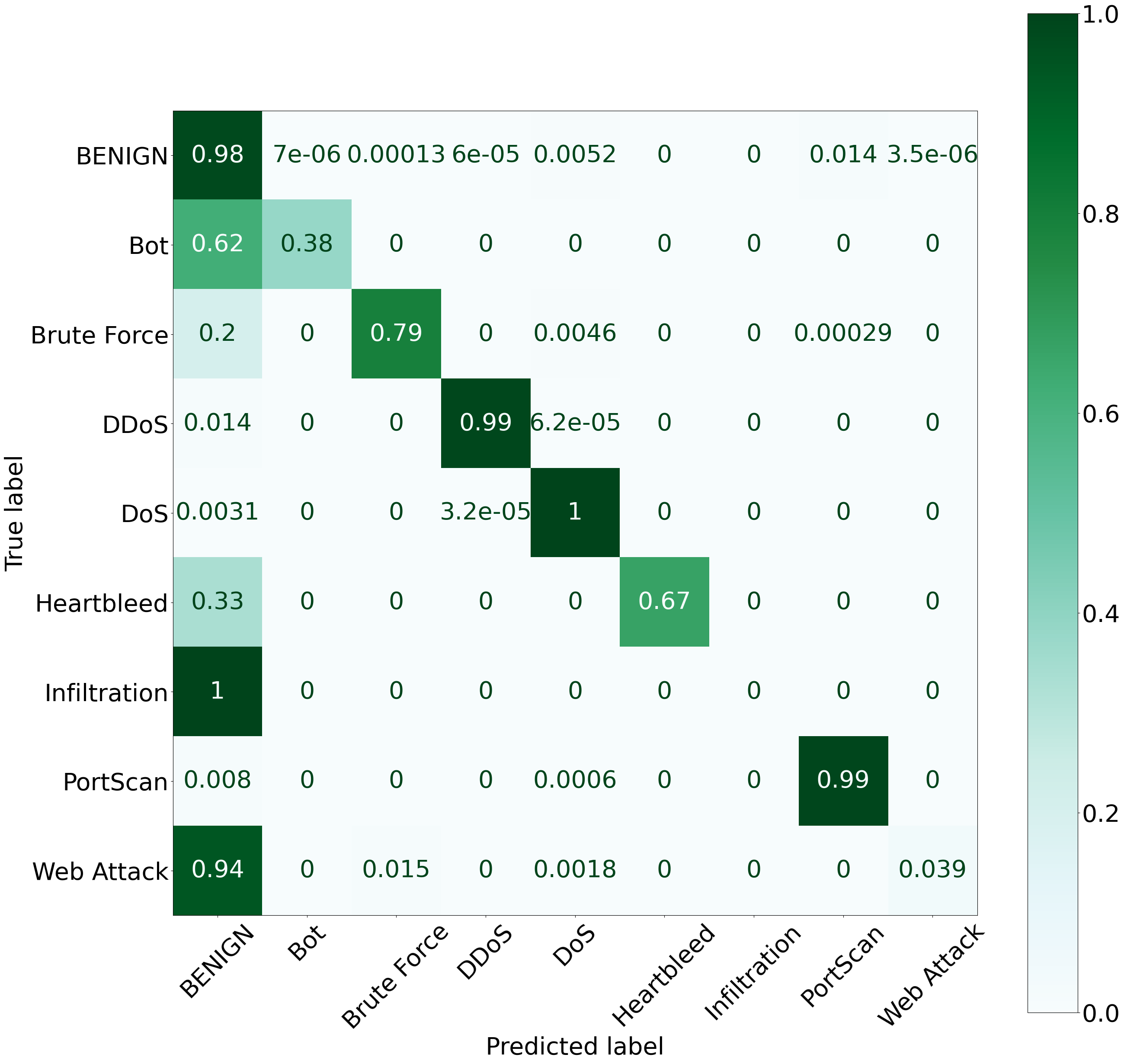}
        \caption{Using sparse categorical cross-entropy loss, no SMOTE.}
        \label{cm_cicids2017_subfig1}
    \end{subfigure}
       \hfill
    \begin{subfigure}{0.84\linewidth} 
        \includegraphics[width=\linewidth]{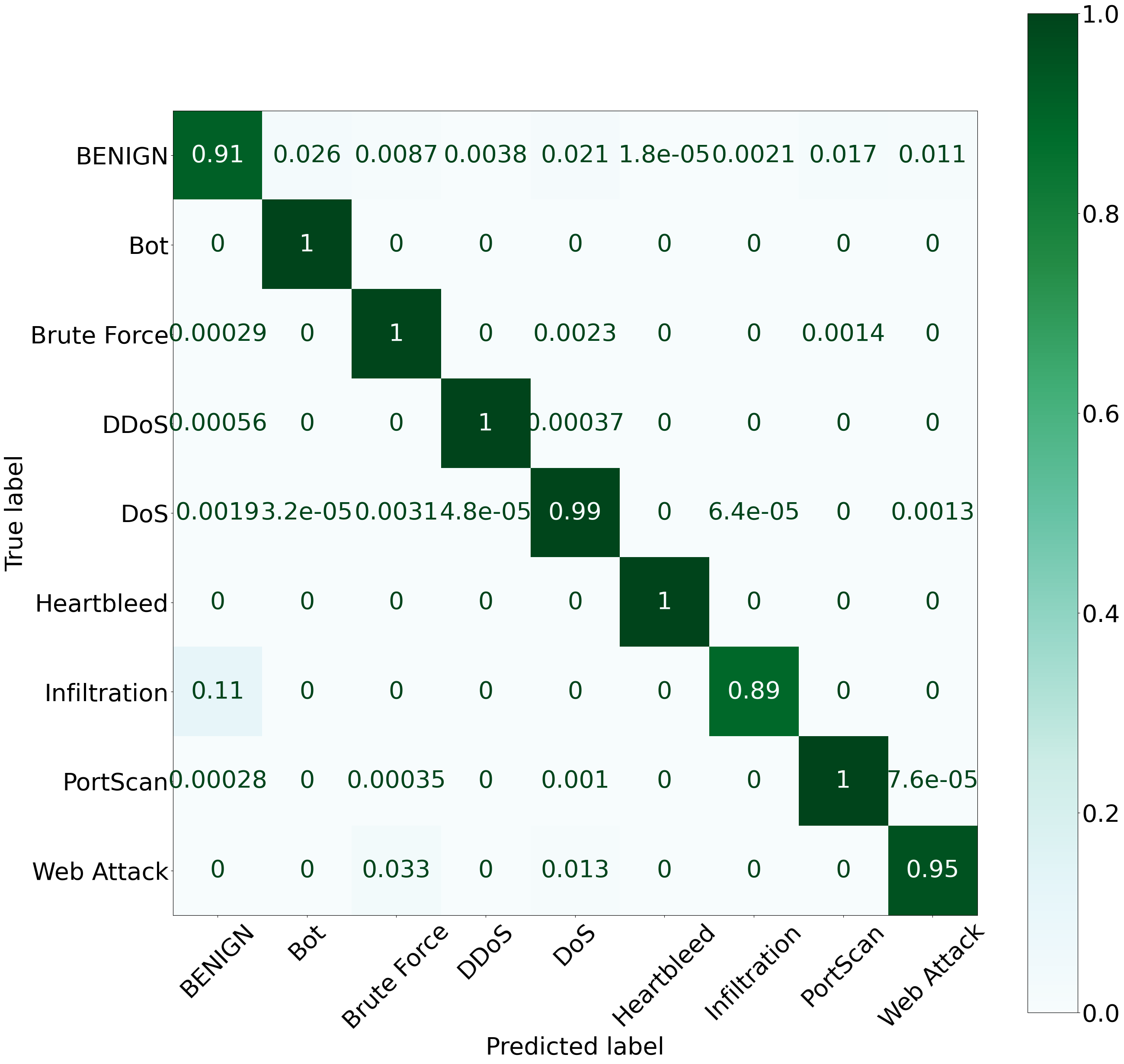}
        \caption{Using sparse categorical cross-entropy loss with SMOTE sampling.}
        \label{cm_cicids2017_subfig2}
    \end{subfigure}
    \hfill
    \begin{subfigure}{0.84\linewidth} 
        \includegraphics[width=\linewidth]{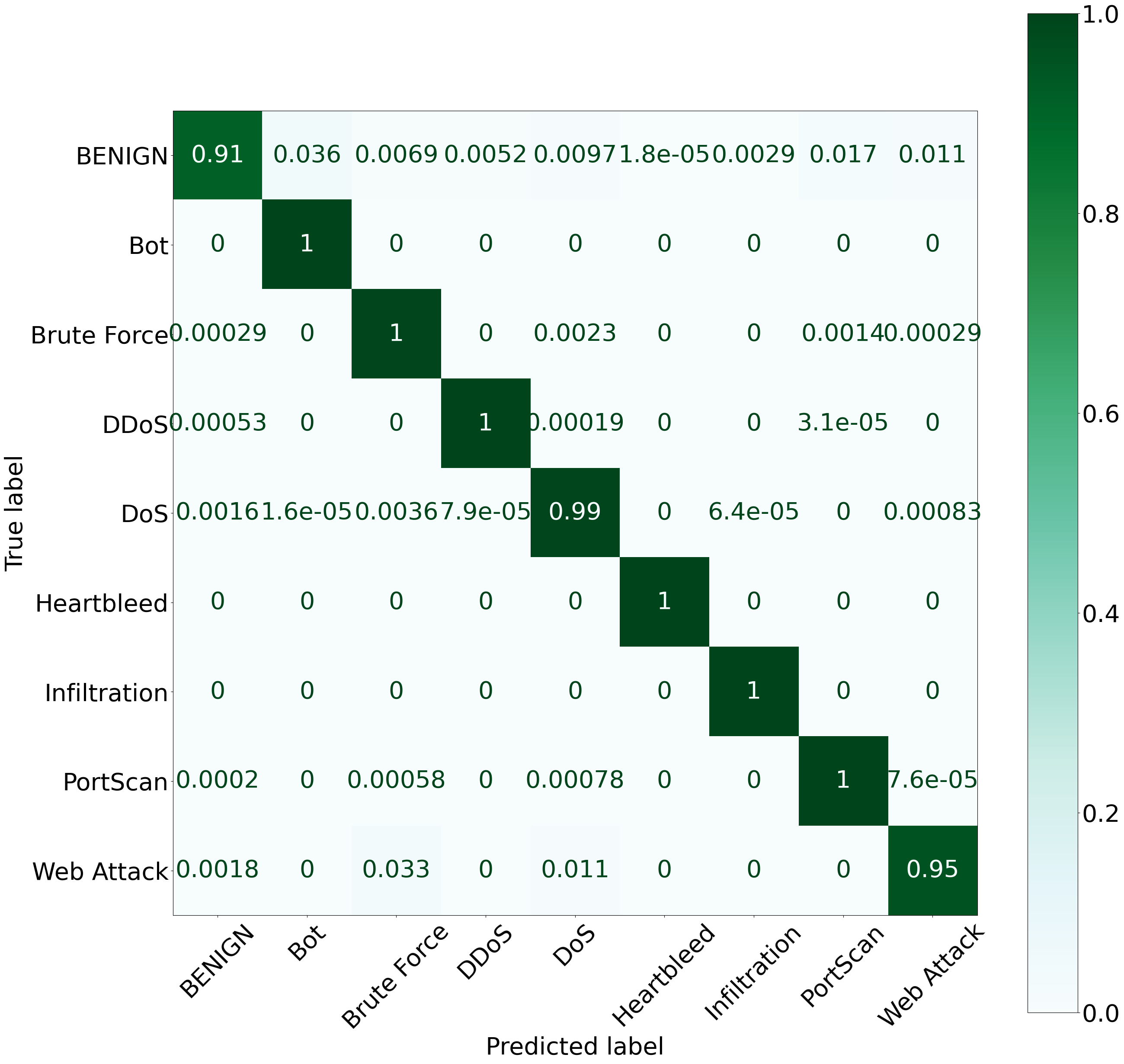}
        \caption{Using categorical focal cross-entropy loss with SMOTE sampling.}
        \label{cm_cicids2017_subfig3} 
    \end{subfigure}
    \caption{Confusion matrices for CICIDS2017 test dataset.}
    \label{cm_cicids2017_fig}
\end{figure}

Figure \ref{cm_cicids2017_fig} shows the confusion matrices computed using the CICIDS2017 dataset. The matrix in \ref{cm_cicids2017_subfig1} shows a failure to detect correct class of network intrusions. The matrix in \ref{cm_cicids2017_subfig3} indicates superior performance over the matrix in \ref{cm_cicids2017_subfig2}. This improvement in the detection of the minority classes is attributed to the CFCL and the SMOTE sampling along with the LSTM architecture.  


Table \ref{train_test_cicids2017_table} shows the comparison of the proposed method on the training and test sets for the CICIDS2017 dataset.
\begin{table*}[t]
\centering
\caption{Comparison of the training and test results on the CICIDS2017 dataset by using the proposed method with SMOTE sampling and focal loss.}\label{train_test_cicids2017_table}
\resizebox{0.55\textwidth}{!}{%
\begin{tabular}{|c|c|c|c|c|c|c|c|c|c|c|}
\hline
\multirow{2}{*}{\textbf{Metric}} & \multicolumn{4}{c|}{\textbf{Training}} & \multicolumn{4}{c|}{\textbf{Test}} \\ \cline{2-9} 
 & \textbf{ACC} & \textbf{PPV} & \textbf{TPR} & \textbf{F1} & \textbf{ACC} & \textbf{PPV} & \textbf{TPR} & \textbf{F1} \\ \hline
\textbf{BENIGN} & 0.99 & 0.96 & 0.99 & 0.97 & 0.97 & 1.00 & 0.91 & 0.95 \\ 
\textbf{Bot} & 1.00 & 1.00 & 0.99 & 0.99 & 0.99 & 0.41 & 0.98 & 0.58 \\ 
\textbf{Brute Force} & 1.00 & 1.00 & 1.00 & 1.00 & 1.00 & 0.85 & 1.00 & 0.92 \\ 
\textbf{DDoS} & 1.00 & 1.00 & 1.00 & 1.00 & 1.00 & 0.98 & 1.00 & 0.99 \\ 
\textbf{DoS} & 1.00 & 1.00 & 0.99 & 0.99 & 1.00 & 0.96 & 1.00 & 0.98 \\ 
\textbf{Heartbleed} & 1.00 & 1.00 & 1.00 & 1.00 & 1.00 & 0.85 & 1.00 & 0.92 \\ 
\textbf{Infiltration} & 1.00 & 1.00 & 1.00 & 1.00 & 1.00 & 0.91 & 1.00 & 0.95 \\ 
\textbf{PortScan} & 1.00 & 1.00 & 0.99 & 0.99 & 0.99 & 0.86 & 1.00 & 0.92 \\ 
\textbf{Web Attack} & 1.00 & 1.00 & 0.99 & 0.99 & 0.99 & 0.92 & 0.95 & 0.93 \\ \hline
\textbf{Average} & 1.00 & 1.00 & 0.99 & 0.99 & 0.99 & 0.86 & 0.98 & 0.90 \\ \hline
\end{tabular}
}
\end{table*}
The proposed method utilizing SMOTE sampling and focal loss showed impressive performance during training, achieving a high average accuracy of 1.00 and strong precision and recall values. However, during testing, the model maintained a relatively high average accuracy (0.99) but faced challenges in correctly identifying certain attack categories, notably Web Attack (PPV: 0.92, TPR: 0.95, F1: 0.93). 

Table \ref{comparison_cicids2017_table} shows the comparison of the proposed method with the LSTM-based method \cite{boukhalfa2020lstm} and the method that utilizes both convolutional layers and the LSTM layers \cite{sun2020dl}.
\begin{table}[ht]
\centering
\caption{Comparison of evaluation metrics given in percentage for different methods on CICIDS2017 dataset.} \label{comparison_cicids2017_table}
\resizebox{0.45\textwidth}{!}{%
\begin{tabular}{|m{2cm}|c|c|c|c|}
\hline
\textbf{Method} & \textbf{ACC} & \textbf{PPV} & \textbf{TPR} & \textbf{F1} \\ \hline
MultinomialNB & 72.52 & 39.56 & 78.20 & 52.54 \\ 
Random Forest & 96.08 & 69.89 & 85.47 & 76.90 \\ 
J48 & 97.32 & 80.56 & 92.23 & 86.02 \\ 
Logistic Regression & 97.68 & 72.56 & 84.96 & 78.27 \\ 
LSTM \cite{boukhalfa2020lstm} & 96.83 & 74.23 & 94.21 & 83.04 \\ 
CNN+LSTM \cite{sun2020dl} & 98.67 & \textbf{87.79} & 97.21 & \textbf{92.26} \\ 
LSTM+SMOTE (Proposed) & \textbf{99.33} & 86.43 & \textbf{98.22} & 91.95 \\ \hline
\end{tabular}
}
\end{table}
The results indicate that the MultinomialNB model achieves the lowest accuracy among the methods, along with relatively lower precision and F1 score. This indicates a weaker overall performance, likely due to its underlying assumptions not aligning well with the dataset's characteristics. In contrast, the Random Forest and J48 (C4.5) models demonstrate significantly higher accuracy, precision, recall, and F1 score, showcasing their effectiveness in handling this dataset. Their decision tree-based nature allows them to capture complex relationships. The Logistic Regression model performs similarly well, indicating the suitability of linear models for this classification task. It achieves high accuracy, precision, recall, and F1 score, suggesting a good balance between true positives and false positives/negatives. The LSTM model, as reported by Boukhalfa et al. \cite{boukhalfa2020lstm}, showcases strong performance, particularly in recall and F1 score. This indicates its ability to effectively identify true positives and maintain a good balance between precision and recall. The CNN+LSTM hybrid model, as proposed by Sun et al. \cite{sun2020dl}, demonstrates outstanding accuracy, precision, recall, and F1 score. The utilization of convolutional and recurrent layers appears to capture complex features and temporal dependencies effectively. However, the use of multiple convolutional and recurrent layers makes the training of their model slower as compared to the proposed method. Finally, the proposed LSTM+SMOTE method outperforms in accuracy and recall, showcasing its strength in identifying true positives. However, it has a slightly lower precision compared to some other models, indicating a trade-off between precision and recall.
\section{Conclusion}
\label{conc}
This paper presented an LSTM-based deep learning architecture to detect various network attacks as a multi-class problem. The use of the SMOTE sampling and focal loss enhanced the network's ability to detect attacks corresponding to the monitory classes in the datasets. Experiments on the CICIDS2017 and KDD99 datasets demonstrated the generalization capabilities of the proposed method for network intrusion detection. 

\footnotesize{
\bibliographystyle{IEEEtran}
\bibliography{main_Rashid}

\begin{thebibliography}{10}
\providecommand{\url}[1]{#1}
\csname url@samestyle\endcsname
\providecommand{\newblock}{\relax}
\providecommand{\bibinfo}[2]{#2}
\providecommand{\BIBentrySTDinterwordspacing}{\spaceskip=0pt\relax}
\providecommand{\BIBentryALTinterwordstretchfactor}{4}
\providecommand{\BIBentryALTinterwordspacing}{\spaceskip=\fontdimen2\font plus
\BIBentryALTinterwordstretchfactor\fontdimen3\font minus
  \fontdimen4\font\relax}
\providecommand{\BIBforeignlanguage}[2]{{%
\expandafter\ifx\csname l@#1\endcsname\relax
\typeout{** WARNING: IEEEtran.bst: No hyphenation pattern has been}%
\typeout{** loaded for the language `#1'. Using the pattern for}%
\typeout{** the default language instead.}%
\else
\language=\csname l@#1\endcsname
\fi
#2}}
\providecommand{\BIBdecl}{\relax}
\BIBdecl

\bibitem{lee2014network}
S.~Lee, K.~Levanti, and H.~S. Kim, ``Network monitoring: Present and future,''
  \emph{Computer Networks}, vol.~65, pp. 84--98, 2014.

\bibitem{alrawashdeh2016toward}
K.~Alrawashdeh and C.~Purdy, ``Toward an online anomaly intrusion detection
  system based on deep learning,'' in \emph{2016 15th IEEE international
  conference on machine learning and applications (ICMLA)}.\hskip 1em plus
  0.5em minus 0.4em\relax IEEE, 2016, pp. 195--200.

\bibitem{azab2022network}
A.~Azab, M.~Khasawneh, S.~Alrabaee, K.-K.~R. Choo, and M.~Sarsour, ``Network
  traffic classification: Techniques, datasets, and challenges,'' \emph{Digital
  Communications and Networks}, 2022.

\bibitem{sun2020dl}
P.~Sun, P.~Liu, Q.~Li, C.~Liu, X.~Lu, R.~Hao, and J.~Chen, ``Dl-ids: Extracting
  features using cnn-lstm hybrid network for intrusion detection system,''
  \emph{Security and Communication Networks}, 2020.

\bibitem{barabas2011evaluation}
M.~Barabas, G.~Boanea, A.~B. Rus, V.~Dobrota, and J.~Domingo-Pascual,
  ``Evaluation of network traffic prediction based on neural networks with
  multi-task learning and multiresolution decomposition,'' in \emph{2011 IEEE
  7th International Conference on Intelligent Computer Communication and
  Processing}.\hskip 1em plus 0.5em minus 0.4em\relax IEEE, 2011, pp. 95--102.

\bibitem{kdd99_dataset}
G.~Srivastava, R.~Cooley, M.~Deshpande, and P.-N. Tan, ``Kdd cup 1999 data,''
  \url{https://archive.ics.uci.edu/dataset/130/kdd+cup+1999+data}, 1999.

\bibitem{tavallaee2009detailed}
M.~Tavallaee, E.~Bagheri, and W.~Lu, ``A detailed analysis of the kdd cup 99
  data set,'' in \emph{2009 International Conference on Computational
  Intelligence for Security and Defense Applications}, 2009, pp. 53--58.

\bibitem{sharafaldin2018toward}
I.~Sharafaldin, A.~H. Lashkari, and A.~A. Ghorbani, ``Toward generating a new
  intrusion detection dataset and intrusion traffic characterization,''
  \emph{ICISSp}, vol.~1, pp. 108--116, 2018.

\bibitem{panigrahi2018detailed}
R.~Panigrahi and S.~Borah, ``A detailed analysis of cicids2017 dataset for
  designing intrusion detection systems,'' \emph{International Journal of
  Engineering \& Technology}, vol.~7, no. 3.24, pp. 479--482, 2018.

\bibitem{fries2008fuzzy}
T.~P. Fries, ``A fuzzy-genetic approach to network intrusion detection,'' in
  \emph{Proceedings of the 10th Annual Conference Companion on Genetic and
  Evolutionary Computation}, 2008, pp. 2141--2146.

\bibitem{koc2012network}
L.~Koc, T.~A. Mazzuchi, and S.~Sarkani, ``A network intrusion detection system
  based on a hidden naïve bayes multiclass classifier,'' \emph{Expert Systems
  with Applications}, vol.~39, no.~18, pp. 13\,492--13\,500, 2012.

\bibitem{zanero2004unsupervised}
S.~Zanero and S.~M. Savaresi, ``Unsupervised learning techniques for an
  intrusion detection system,'' in \emph{Proceedings of the 2004 ACM symposium
  on Applied computing}.\hskip 1em plus 0.5em minus 0.4em\relax ACM, 2004, pp.
  412--419.

\bibitem{heba2010principal}
F.~E. Heba, A.~Darwish, A.~E. Hassanien, and A.~Abraham, ``Principal components
  analysis and support vector machine based intrusion detection system,'' in
  \emph{2010 10th International Conference on Intelligent Systems Design and
  Applications}.\hskip 1em plus 0.5em minus 0.4em\relax IEEE, 2010, pp.
  363--367.

\bibitem{alalousi2016preliminary}
A.~Alalousi, M.~Razif, M.~AbuAlhaj, M.~Anbar, and S.~Nizam, ``A preliminary
  performance evaluation of k-means, knn and em unsupervised machine learning
  methods for network flow classification,'' \emph{International Journal of
  Electrical and Computer Engineering}, vol.~6, no.~2, p. 778, 2016.

\bibitem{ravale2015feature}
U.~Ravale, N.~Marathe, and P.~Padiya, ``Feature selection based hybrid anomaly
  intrusion detection system using k means and rbf kernel function,''
  \emph{Procedia Computer Science}, vol.~45, pp. 428--435, 2015.

\bibitem{chen2020fuzzy}
L.~Chen, S.~Gao, B.~Liu, Z.~Lu, and Z.~Jiang, ``Few-nnn: A fuzzy entropy
  weighted natural nearest neighbor method for flow-based network traffic
  attack detection,'' \emph{China Communications}, vol.~17, no.~5, pp.
  151--167, 2020.

\bibitem{yu2020intrusion}
Y.~Yu and N.~Bian, ``An intrusion detection method using few-shot learning,''
  \emph{IEEE Access}, vol.~8, pp. 49\,730--49\,740, 2020.

\bibitem{malaiya2018empirical}
R.~K. Malaiya, D.~Kwon, J.~Kim, S.~C. Suh, H.~Kim, and I.~Kim, ``An empirical
  evaluation of deep learning for network anomaly detection,'' in \emph{2018
  International Conference on Computing, Networking and Communications
  (ICNC)}.\hskip 1em plus 0.5em minus 0.4em\relax IEEE, 2018, pp. 893--898.

\bibitem{wang2018network}
W.~Wang, Y.~Bai, C.~Yu, Y.~Gu, P.~Feng, X.~Wang, and R.~Wang, ``A network
  traffic flow prediction with deep learning approach for large-scale
  metropolitan area network,'' in \emph{NOMS 2018-2018 IEEE/IFIP Network
  Operations and Management Symposium}.\hskip 1em plus 0.5em minus 0.4em\relax
  IEEE, 2018, pp. 1--9.

\bibitem{li2018detection}
C.~Li, Y.~Wu, X.~Yuan, Z.~Sun, W.~Wang, X.~Li, and L.~Gong, ``Detection and
  defense of ddos attack--based on deep learning in openflow-based sdn,''
  \emph{International Journal of Communication Systems}, vol.~31, no.~5, p.
  e3497, 2018.

\bibitem{ullah2019cyber}
F.~Ullah, H.~Naeem, S.~Jabbar, S.~Khalid, M.~A. Latif, F.~Al-Turjman, and
  L.~Mostarda, ``Cyber security threats detection in internet of things using
  deep learning approach,'' \emph{IEEE access}, vol.~7, pp. 124\,379--124\,389,
  2019.

\bibitem{jiang2020network}
K.~Jiang, W.~Wang, A.~Wang, and H.~Wu, ``Network intrusion detection combined
  hybrid sampling with deep hierarchical network,'' \emph{IEEE Access}, vol.~8,
  pp. 32\,464--32\,476, 2020.

\bibitem{li2019method}
J.~Li, X.~Yun, M.~Tian, J.~Xie, S.~Li, Y.~Zhang, and Y.~Zhou, ``A method of
  http malicious traffic detection on mobile networks,'' in \emph{2019 IEEE
  Wireless Communications and Networking Conference (WCNC)}.\hskip 1em plus
  0.5em minus 0.4em\relax IEEE, 2019, pp. 1--8.

\bibitem{yan2018effective}
B.~Yan and G.~Han, ``Effective feature extraction via stacked sparse
  autoencoder to improve intrusion detection system,'' \emph{IEEE Access},
  vol.~6, pp. 41\,238--41\,248, 2018.

\bibitem{sun2022detecting}
Y.~Sun, Y.~C. Lu, K.~Fu, F.~Chen, and C.~T. Lu, ``Detecting anomalous traffic
  behaviors with seasonal deep kalman filter graph convolutional neural
  networks,'' \emph{Journal of King Saud University-Computer and Information
  Sciences}, vol.~34, no.~8, pp. 4729--4742, 2022.

\bibitem{hinton2006fast}
G.~E. Hinton, S.~Osindero, and Y.-W. Teh, ``A fast learning algorithm for deep
  belief nets,'' \emph{Neural computation}, vol.~18, no.~7, pp. 1527--1554,
  2006.

\bibitem{marir2018distributed}
N.~Marir, H.~Wang, G.~Feng, B.~Li, and M.~Jia, ``Distributed abnormal behavior
  detection approach based on deep belief network and ensemble svm using
  spark,'' \emph{IEEE Access}, vol.~6, pp. 59\,657--59\,671, 2018.

\bibitem{wei2019optimization}
P.~Wei, Y.~Li, Z.~Zhang, T.~Hu, Z.~Li, and D.~Liu, ``An optimization method for
  intrusion detection classification model based on deep belief network,''
  \emph{IEEE Access}, vol.~7, pp. 87\,593--87\,605, 2019.

\bibitem{yu2021network}
T.~Yu, X.~Yin, M.~Yao, and T.~Liu, ``Network security monitoring method based
  on deep learning,'' in \emph{Journal of Physics: Conference Series}, vol.
  1955, no.~1.\hskip 1em plus 0.5em minus 0.4em\relax IOP Publishing, 2021, p.
  012040.

\bibitem{xu2018intrusion}
C.~Xu, J.~Shen, X.~Du, and F.~Zhang, ``An intrusion detection system using a
  deep neural network with gated recurrent units,'' \emph{IEEE Access}, vol.~6,
  pp. 48\,697--48\,707, 2018.

\bibitem{gwon2019network}
H.~Gwon, C.~Lee, R.~Keum, and H.~Choi, ``Network intrusion detection based on
  lstm and feature embedding,'' \emph{arXiv preprint arXiv:1911.11552}, 2019.

\bibitem{boukhalfa2020lstm}
A.~Boukhalfa, A.~Abdellaoui, N.~Hmina, and H.~Chaoui, ``Lstm deep learning
  method for network intrusion detection system,'' \emph{International Journal
  of Electrical and Computer Engineering}, vol.~10, no.~3, p. 3315, 2020.

\bibitem{chawla2002smote}
N.~V. Chawla, K.~W. Bowyer, L.~O. Hall, and W.~P. Kegelmeyer, ``Smote:
  Synthetic minority over-sampling technique,'' \emph{Journal of Artificial
  Intelligence Research}, vol.~16, pp. 321--357, 06 2002.

\bibitem{hochreiter1997long}
S.~Hochreiter, ``Long short-term memory,'' \emph{Neural Computation}, vol.~9,
  no.~8, pp. 1735--1780, 1997.

\bibitem{imrana2018bidirectional}
Y.~Imrana, Y.~Xiang, L.~Ali, and Z.~Abdul-Rauf, ``A bidirectional lstm deep
  learning approach for intrusion detection,'' \emph{Expert Systems With
  Applications}, vol.~93, pp. 418--429, 2018.

\end{thebibliography}
}

\vfill\break

\end{document}